\RequirePackage{fix-cm}
\documentclass[smallextended]{svjour3}       
\smartqed  
\usepackage{graphicx}
\usepackage{natbib}
\usepackage[bookmarks=true,bookmarksnumbered=true,colorlinks=false,pdfborder={0 0 0},hyperfootnotes=false]{hyperref}
\journalname{Experimental Astronomy}
\begin{document}


\title{OSS (Outer Solar System): A fundamental and planetary physics mission to Neptune, Triton and the Kuiper Belt
}

\titlerunning{OSS (Outer Solar System) Mission}        

\author{B.~Christophe 
\and L.J.~Spilker 
\and J.D.~Anderson
\and N.~Andr\'e
\and S.W.~Asmar
\and J.~Aurnou
\and D.~Banfield
\and A.~Barucci
\and O.~Bertolami
\and R.~Bingham
\and P.~Brown
\and B.~Cecconi
\and J.-M.~Courty
\and H.~Dittus
\and L.N.~Fletcher
\and B.~Foulon
\and F.~Francisco
\and P.J.S.~Gil
\and K.H.~Glassmeier
\and W.~Grundy
\and C.~Hansen
\and J.~Helbert
\and R.~Helled
\and H.~Hussmann
\and B.~Lamine
\and C.~L\"ammerzahl
\and L.~Lamy
\and R.~Lehoucq
\and B.~Lenoir
\and A.~Levy
\and G.~Orton
\and J.~P\'aramos
\and J.~Poncy
\and F.~Postberg
\and S.V.~Progrebenko
\and K.R. Reh
\and S.~Reynaud
\and C.~Robert
\and E.~Samain
\and J.~Saur
\and K.M.~Sayanagi
\and N.~Schmitz
\and H.~Selig
\and F.~Sohl
\and T.R.~Spilker
\and R.~Srama
\and K.~Stephan
\and P.~Touboul
\and P.~Wolf
}


\institute{
B.~Christophe, B.~Foulon, B.~Lenoir, A.~Levy, C.~Robert, P.~Touboul\at ONERA - The French Aerospace Lab, F-92322 Ch\^atillon, France \\
              \email{bruno.christophe(at)onera.fr}           
\and L.J.~Spilker, J.D.~Anderson, S.W.~Asmar, K.R.~Reh, G.~Orton, T.R.~Spilker \at JPL/NASA (USA)
\and R.~Lehoucq \at CEA Saclay, Service d'Astrophysique (France)
\and N.~Andr\'e \at IRAP, CNRS, Univ. Paul Sabatier Toulouse (France)
\and D.~Banfield \at Cornell University (USA)
\and J.~Helbert, H.~Hussmann, N.~Schmitz, F.~Sohl, K. Stephan \at DLR/Institute of Planetary Research (Germany)
\and H.~Dittus \at DLR/Institute of Space System (Germany)
\and P.~Brown \at Imperial College London (UK)
\and R.~Srama \at IRS, University of Stuttgart and MPIK, Heidelberg (Germany)
\and F.~Francisco, P.J.S.~Gil, J.~Paramos \at Instituto Superior T\'ecnico, Universidade T\'ecnica de Lisboa (Portugal)
\and S.V.~Progrebenko \at JIVE, Joint Institute for VLBI in Europe (The Netherlands)
\and A.~Barucci, B.~Cecconi, L.~Lamy \at Laboratoire d'Etudes Spatiales et d'Instrumentation en Astrophysique, Observatoire de Paris, CNRS, Univ. Pierre et Marie Curie, Univ. Paris Diderot, F-92195 Meudon (France)
\and J.-M.~Courty, B.~Lamine, S.~Reynaud \at LKB, CNRS, Paris (France)
\and W.~Grundy \at Lowel Observatory (USA)
\and E.~Samain \at Observatoire de la C\^ote d'Azur, GeoAzur (France)
\and C.~Hansen \at PSI (USA)
\and R.~Bingham \at RAL (UK)
\and P.~Wolf \at LNE-SYRTE, Observatoire de Paris, CNRS, UPMC (France)
\and J.~Poncy \at Thales Alenia Space, Cannes (France)
\and K.H.~Glassmeier \at Technical University of Braunschweig (Germany)
\and O.~Bertolami \at Universidade do Porto (Portugal)
\and J.~Saur \at Universit\"at zu K\"oln (Germany)
\and J.~Aurnou, R.~Helled \at University of California Los Angeles (USA)
\and K.M.~Sayanagi \at Hampton University in Virginia (USA)
\and F.~Postberg \at University of Heidelberg (Germany)
\and L.N.~Fletcher \at University of Oxford (UK)
\and C.~L\"ammerzahl, H.~Selig \at ZARM, University of Bremen (Germany)
}

\date{Received: date / Accepted: date}

\maketitle


\begin{abstract}

The present OSS (Outer Solar System) mission continues a long and bright tradition by associating the communities of fundamental physics and planetary sciences in a single mission with ambitious goals in both domains. OSS is an M-class mission to explore the Neptune system almost half a century after the flyby of the Voyager 2 spacecraft.

Several discoveries were made by Voyager~2, including the Great Dark Spot (which has now disappeared) and Triton's geysers. Voyager~2 revealed the dynamics of Neptune's atmosphere and found four rings and evidence of ring arcs above Neptune. 
Benefiting from a greatly improved instrumentation, a mission as OSS would result in a striking advance in the study of the farthest planet of the solar system. 
Furthermore, OSS would provide a unique opportunity to visit a selected Kuiper Belt object subsequent to the passage of the Neptunian system. OSS would help consolidate the hypothesis of the origin of Triton as a Kuiper Belt object captured by Neptune, and to improve our knowledge on the formation of the solar system. 

The OSS probe would carry instruments allowing precise tracking of the spacecraft
during the cruise. It would facilitate the best possible tests of the
laws of gravity in deep space. These objectives are important for
fundamental physics, as they test General Relativity, our current
theoretical description of gravitation, but also for cosmology,
astrophysics and planetary science, as General Relativity is used as
a tool in all these domains. In particular, the models of solar
system formation uses General Relativity to describe the crucial
role of gravity.

OSS is proposed as an international cooperation between ESA and NASA, giving the capability for ESA to launch an M-class mission towards the farthest planet of the solar system, and to a Kuiper Belt object. The proposed mission profile would allow to deliver a 500 kg class spacecraft. The design of the probe is mainly constrained by the deep space gravity test in order to minimize the perturbation of the accelerometer measurement.

\end{abstract}

\section{Introduction}\label{section:introduction}

Gravitational physics and solar system physics have been intimately connected at various stages of their developments. 
Isaac Newton used the movement of planets as a crucial laboratory for testing his new theory.
Pierre-Simon Laplace extended the tools of mathematical physics of gravity and developed the nebular hypothesis of the origin of the solar system. 
Urbain Jean Joseph Le Verrier calculated the position of the "eighth planet" by analyzing the perturbations of the orbit of Uranus. 
Johann Gottfried Galle discovered the new planet, later on to be named Neptune, at the predicted position. 
Le Verrier also analyzed the anomaly of the motion of Mercury which, after a lot of discussions between astronomers and physicists, became the first proof of the validity of general relativity \citep{Earman-1993}.
\newline
The present Outer Solar System Mission (OSS) proposed in the frame of a M mission continues this long and bright tradition by associating the communities of fundamental physics and planetary sciences in a mission with ambitious goals in both domains. 
OSS would visit Neptune and its moon Triton, nearly half a century after Voyager~2. 
Using a suite of advanced instrumentation with strong heritage from previous outer solar system missions, OSS would provide striking advances in the study of the farthest known planet of the solar system. 
The Neptune flyby would be precisely controlled to permit a close encounter with a Kuiper Belt object (KBO) to be properly chosen among the large number of scientifically interesting and attainable objects (owing to the large mass of Neptune, this number being much larger than for New Horizons, NASA's fast-track mission towards Pluto and a KBO). 
A mission like OSS would have the potential to consolidate the hypothesis of the origin of Triton as a KBO captured by Neptune at the time of formation of the solar system. 
\newline
The OSS probe would carry instruments allowing a precise tracking of the spacecraft during the cruise. It will make possible the best ever tests of the laws of
gravity in the outer solar system. General Relativity will be tested
in deep space with an unprecedented accuracy, more than a hundred
times better than currently done. This is important not only for
fundamental physics, but also for cosmology and astrophysics in a
context where the observations currently interpreted in terms of
dark matter and dark energy, challenge General Relativity at scales
much larger than that of the solar system. The scientific goal of
better tests of the gravity laws is also directly connected to the
question of the origin of the solar system as models of solar system
formation uses General Relativity to describe the crucial role of
gravity. Using laser metrology, the OSS mission will also improve
the result of the Cassini spacecraft that measured Eddington's
parameter $\gamma$ during its interplanetary journey to Saturn.
\newline
After a brief description of the scientific objectives of the missions, the instrumentation suite is presented. 
A brief analysis of the mission profile is performed. 
Then the spacecraft design is described.

\section{Scientific objectives}\label{section:objectives}

\subsection{Fundamental Physics}\label{subsection:fundamental}

\subsubsection{Deep space gravity}\label{subsubsection:deep-space-gravity}

General Relativity, the current theoretical formulation of gravitation, is in good agreement with most experimental tests of gravitation \citep{Will-2006}. But General Relativity is a classical theory and all attempts to merge it with the quantum description of the other fundamental interactions suggest that it cannot be the final theory of gravitation. Meanwhile, the experimental tests leave open windows for deviations from General Relativity at short \citep{Adelberger-2003} or long distance \citep{Jaekel-2005} scales.

General Relativity is also challenged by observations at galactic
and cosmic scales. The rotation curves of galaxies and the relation
between redshifts and luminosities of supernovae deviate from the
predictions of the theory. These anomalies are interpreted as
revealing the presence of new components of the Universe, the
so-called "dark matter" and "dark energy" \citep{Copeland-2006,
Frieman-2008} which are thought to constitute respectively 23\% and
72\% of the energy content of the Universe. Their nature remains
unknown and, despite their prevalence, they have not been detected
by any other means than gravitational measurements. Given the
immense challenge posed by these large scale behaviors, it is
important to explore every possible explanation including the
hypothesis that General Relativity is not a correct description of
gravity at large scales \citep{Aguirre-2001, Nojiri-2007}.

Testing gravity at the largest scales reachable by man-made instruments is therefore essential to bridge the gap between experiments in the solar system and astrophysical or cosmological observations. The most notable existing test in this domain was performed by NASA during the extended Pioneer 10 \& 11 missions. This test resulted in what is now known as the Pioneer anomaly \citep{Anderson-1998, Anderson-2002}, one of the few experimental signals deviating from the predictions of General Relativity \citep{Laemmerzahl-2008, Anderson-2009}.

In a context dominated by the quest for the nature of dark matter and dark energy, the challenge raised by the anomalous Pioneer signals has to be faced. Efforts have been devoted to the reanalysis of Pioneer data \citep{Markwardt-2002, Olsen-2007, Bertolami-2008, Levy-2009, Turyshev-2009}), with the aim of learning as much as possible on its possible origin, which can be an experimental artifact \citep{Bertolami-2012, Rievers-2011, Turyshev-2012} as well as a hint of considerable importance for fundamental physics \citep{Turyshev-2010}. In the meantime, theoretical studies have been devoted to determine whether or not the anomalous signal could reveal a scale-dependent modification of the law of gravity while remaining compatible with other tests. Among the candidates, one finds metric extensions \citep{Jaekel-2005, Jaekel-2005b, Jaekel-2006, Jaekel-2006b} as well as field theoretical models \citep{Bertolami-2004, Moffat-2005, Moffat-2006, Brownstein-2006, Bruneton-2007, Bertolami-2007} of General Relativity.

Several mission concepts have been put forward \citep{Anderson-2002,
Dittus-2005, Johann-2007, Bertolami-2007, Christophe-2009,
Wolf-2009} to improve the experiment performed by Pioneer 10 \& 11
probes. A key idea in these proposals is to measure
non-gravitational forces acting on the spacecraft, whatever may be
their underlying cause, and thus remove as fully as possible all the ambiguity introduced by systematic effects.
The addition of an accelerometer on
board the spacecraft not only improves the precision and quality of
the navigation but also allows for understanding the origin of any
anomalous signals. The target accuracy of the accelerometry
measurement is 1~pm/s$^{2}$ rms after an integration time of 3
hours. Combining these measurements with radio tracking data, it
becomes possible to improve by 3 orders of magnitude the precision
of the comparison with theory of the spacecraft gravitational
acceleration. 
The deep space measurement will occur several times per year (in order to detect deviation of the general relativity at 0.5 or 1~year) and some long period of one month (in order to detect variation of the general relativity at 0.5 or 1~sidereal day).

The same instrument also improves the science return with respect to objectives in exploration of the outer solar system physics, which is the motivation for combining fundamental physics and planetary physics in a common mission. This idea has been included in the Roadmap for Fundamental Physics in Space issued in 2010 by ESA\footnote{ESA Fundamental Physics Roadmap Advisory Team, A Roadmap for Fundamental Physics in Space, 2010. Available at [08/23/2010]: http://sci.esa.int/fprat}. Let us emphasize at this point that these scientific goals are intimately connected since the law of gravity is directly connected to the planetary ephemeris \citep{Fienga-2010} as well as to the origins of the solar system \citep{Blanc-2005}.

\subsubsection{Measurement of the Eddington's parameter $\gamma$}\label{subsubsection:gamma}

Metric extensions of General Relativity are often characterized in
terms of Eddington parameters $\beta$ and $\gamma$ which measure
deviations from General Relativity \citep{Will-2006}. In particular,
the factor $(1-\gamma)$ gauges the fractional strength of scalar
interaction in scalar-tensor theories of gravity. This deviation
$(1-\gamma)$ has been shown to be smaller than $2\times 10^{-5}$ by
the Cassini relativity experiment performed at solar conjunctions in
June 2002 \citep{Bertotti-2003}. But recent theoretical proposals
suggest that this deviation might have a natural value in the range
$10^{-6}$-$10^{-7}$ as a consequence of a damping of the scalar
contribution to gravity during cosmological evolution
\citep{Damour-2002}.
 
The orbit of the spacecraft will be tracked during the whole cruise phase, in order to test General Relativity to an unprecedented level of accuracy (see previous section).
A particularly interesting test will take benefit of solar conjunctions to repeat the Cassini relativity experiment, which has given the best constraints on deviations from GR to date. 
Doppler observable of the radio or optical link when close to conjunction can be written as:
\begin{equation}
\frac{\Delta \nu}{\nu }=-4 (1+\gamma )\frac{GM}{c^3 b}\frac{db}{dt}
\end{equation}
where $b$ is the impact parameter (distance of closest approach) of the laser beam. 
Supposing observations down to 1.5 solar radii ($b \approx 11 \times 10^8$ m) and approximating $db/dt \approx 30$ km/s (Earth orbital velocity) the maximum effect is about $1 \times 10^{-9}$. 

If using the same radio-science as Cassini (Doppler accuracy $\approx 10^{-13}$ over one day), the experiment will confirm its results with the advantage of an accelerometer on board to measure the non-geodesic acceleration.
A largely improved accuracy can be attained with the up-scaling option of a laser ranging equipment onboard. The OSS mission can thus measure the parameter $(1-\gamma)$ at the $10^{-7}$~level, which would provide new crucial information on scalar-tensor theories of gravity at their fascinating interface with theories of cosmological evolution.

\subsection{Neptune}\label{subsection:neptune}

Extrasolar planet hunting has matured to the point of not only detecting ice-giant-sized bodies around other stars, 
but even measuring the bulk compositional properties and mapping out the spatial characteristics and thermal parameters of these extrasolar planets (e.g., \cite{Harrington-2006}).
Our understanding of these extrasolar ice-giants is hampered by our limited knowledge of many basic aspects of our own nearby ice giants, which should serve as templates for their extrasolar cousins.

\subsubsection{Neptune Interior}\label{subsubsection:neptune-interior}

The interior of Neptune is poorly understood but likely composed of a mixture of rock and ices \citep{Hubbard-1991, Podolak-1995}. It is not clear, however, if rock and ice components are fully or incompletely separated so that density would increase more gradually toward the centre. The radial extent of the core region could amount up to 70\% of the total radius, thereby substantially affecting the planet's low-degree gravitational field. Unfortunately, however, present-day observational constraints on Neptune's interior structure are limited to gravitational harmonics to forth degree (J$_2$, J$_4$) within relatively broad error margins. Whereas substantial thermal excess emission implies adiabaticity of Neptune's deep interior, the presence of a multipolar magnetic field requires electrically conducting fluid regions (probably salty H$_2$O) at shallow depths \citep{Ness-1989, Stanley-2004}.

Neptune's shape and rotational state are still imperfectly known to constrain interior structure models \citep{Helled-2010}. However, the shape of a giant planet contains important constraints on its rotation rate, and can be used to discriminate between different rotation profiles and provides information on the dynamics. The rotation rate of the planet is required for internal modeling (e.g., \citet{Zharkov-1978}). In case that the planet rotates differentially, or if the zonal winds are deep enough, the planetary shape is adjusted accordingly, and corrections to the gravitational coefficients and, therefore, the planetary internal structure must be included in the models \citep{Hubbard-1999, Hubbard-1991}. The planetary shape can also be used to constrain the depth of the zonal winds that are crucial for our understanding of magnetic field generation and global circulation in the planet.

\subsubsection{Neptune's Atmosphere}\label{subsubsection:neptune-atmosphere}

Given its great distance from the sun, Neptune has a surprisingly dynamic atmosphere,
including a jet stream blowing at almost 500~m/s \citep{Limaye_Sromovsky_1991} and a giant
vortex \citep{Smith-1989}.  The great question is how this dynamical weather system is
powered.  Hazes and clouds in the troposphere and stratosphere probably play a major role in
modulating solar heating, which ultimately controls the meridional and vertical profiles of
temperature and winds.  How this thermal energy is converted to kinetic energy remains
unknown.  Studies of jetstreams on Jupiter and Saturn have revealed that small-scale eddies
can provide the momentum forcing necessary to drive the jets (e.g.,
\citet{Salyk_Ingersoll_etal_2006, DelGenio_etal_2007, Aurnou-2007,
Sayanagi_Showman_Dowling_2008}), but - despite Voyager-2 and Earth-based studies showing
rapid variability in large-scale eddies \citep{Luszcz-Cook_etal_2010}, the small eddies that
might feed energy and momentum to the larger weather phenomena are yet to be seen on
Neptune. Similarly, even though multiple generations of Great Dark spots have been observed,
the smaller-scale eddies that may contribute to their maintenance and generation have never
been seen.

High-resolution observations with a camera optimized for Neptune's atmosphere
will enable a search for eddies at the relevant spatial scales.  Similarly measurements of
Neptune's thermal emission from the mid-infrared through the submillimeter can determine the
depth to which differences between both axisymmetric and discrete regions exist.  
Measurements of temperatures and cloud properties, along with the distribution of trace
species and the para- to ortho-H$_2$ ratio will also provide indirect tracers of vertical
winds.  These are essential measurements to determine what powers Neptune's circulation and
how different they and Neptune's thermal structure is from those of Uranus, whose internal
heat source is immeasurably low - in direct contrast to Neptune's. 

\subsubsection{Neptune's magnetic field and magnetosphere}\label{subsubsection:neptune-magnetosphere}

Neptune's magnetic dipole, like that of Uranus, is highly tilted and offset from the
planet's centre \citep{Ness-1989, Ness-1994, Connerney-1991}. The equatorial surface field
is 1.42~$\mu$T, corresponding to a magnetic moment about 27~times greater than at Earth. 
Neptune's large quadrupole moment makes a greater contribution to the surface
magnetic field than at any other planet, which is symptomatic of a very irregular magnetic
field. The octupole and higher moments are essentially undetermined \citep{Connerney-1991}. 
Although \citet{Stanley-2004} have attributed the large tilt and strong quadrupole moment
to a thin shell structure and relatively poor electrical conductivity of the ice mantle
where the magnetic field is thought to be generated, it is inconsistent with a picture 
\citep{Fortney-2011} where convection occurs throughout the fluid envelope.

Neptune's magnetic field goes through dramatic changes as the planet rotates in the solar wind
\citep{Bagenal-1992}, with the magnetosphere being completely reconfigured twice per
planetary rotation period. Thus, it is not clear why, despite this, the magnetosphere
appeared very quiescent during the Voyager 2 flyby in 1989.  Additional observations that
map more fully in time and space than those from the single Voyager-2 flyby will answer
these questions. In contrast with near-solstice observations of Voyager 2, near-equinox
conditions will prevail for approximately 2 decades around 2038, and magnetic
reconnection will be far more favoured (once per rotation) than in 1989 - allowing 
observations of the magnetospheric response to solar wind input on time scales of hours. 
Such information will reveal how magnetospheres work, and can not be obtained by studying
Earth, Jupiter or Saturn alone.  

The plasma in Neptune's magnetosphere is thought to be
derived mainly from Triton \citep{Richardson-1991}.  Auroral emissions, associated with
energetic electrons circulating along high-latitude magnetic field lines, were
unambiguously observed at radio wavelengths (Neptune's Kilometric Radiation - NKR - consists
of no less than 4 different radio components) around north (N) and south (S) magnetic poles
as well as close to the magnetic equator, which is a characteristic of ice giants
\citep{Zarka-1995}.  Atmospheric aurorae were also tentatively observed in UV around the
S~pole \citep{Sandel-1990}. A strong confinement of the radiation belts by the minimum L
shell of Triton was observed \citep{Mauk-1995}; none of these features have been
satisfactorily explained.

\subsection{Triton}\label{subsection:triton}

Triton is now locked in a circular synchronous orbit that is retrograde and highly inclined ($\sim 23^{\circ}$), thereby suggesting an origin by capture from the inner Kuiper Belt. 
Tidal energy release during orbit circularization may have kept Triton's interior sufficiently warm to separate ice and rock from each other. 
While Triton's mean density ($2059 \pm 5$ kg~m$^3$) \citep{Jacobson-2009, Thomas-2000} is consistent with a large rocky core overlain by a relatively thin icy shell, models of the radial mass distribution would be best constrained by radio science measurements of the non spherical part of the satellite's low-degree gravity field. 
As the latter is dominated by both spin and tidal contributions, a level-surface theory as developed by \citet{Hubbard-1978}, \citet{Dermott-1979}, and more recently in a series of papers by \citet{Zharkov-2010} can be used to deduce the concentration of mass toward the center of a synchronously rotating satellite. 

Since the Voyager encounters, subsurface water oceans have been detected in the icy Jovian moons \citep{Zimmer-2000, Kivelson-2002, Schubert-2004} and are predicted in Enceladus and Titan \citep{Schubert-2007, Lorenz08}, as well as in Triton \citep{Stevenson-2002, Hussmann-2006}, indicating that liquid reservoirs may be common in icy moons. 
Determining whether Triton has an ocean, and whether any of its chemistry is expressed on its surface would make Triton an attractive astrobiological target. 
As there are two distinct time-variable magnetic field components in the vicinity of Triton, the existence of a putative subsurface ocean could be inferred from induced magnetic field measurements. 
One is due to the highly tilted magnetic field of Neptune and the other one is due to Triton's large inclination. 
The two field components might allow to resolve the thickness and the conductivity of the ocean separately \citep{Saur-2010}.

\begin{figure}
 \begin{tabular}{cc}
  \includegraphics[width = 0.45 \linewidth]{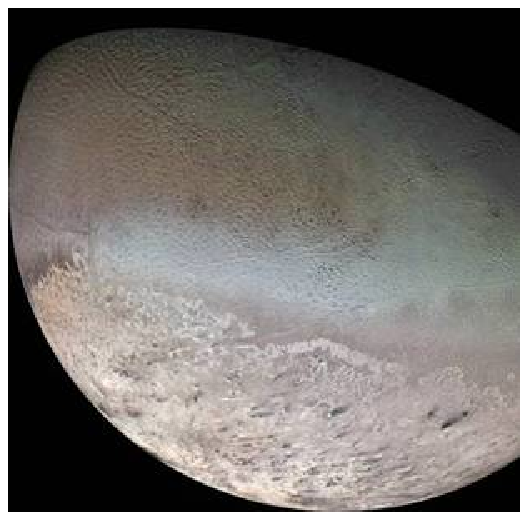} &
  \includegraphics[width = 0.45 \linewidth]{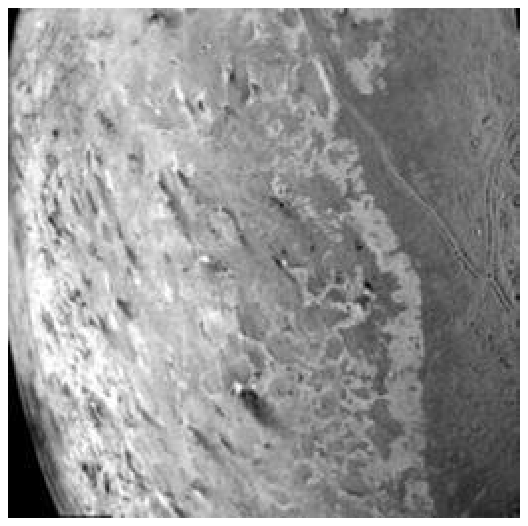} \\
  \includegraphics[width = 0.45 \linewidth]{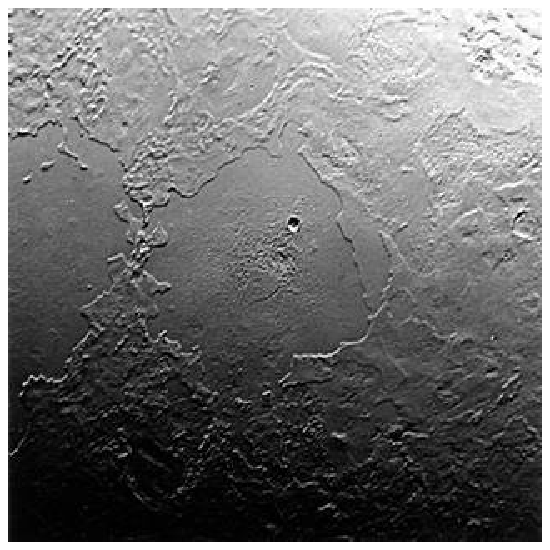} &
  \includegraphics[width = 0.45 \linewidth]{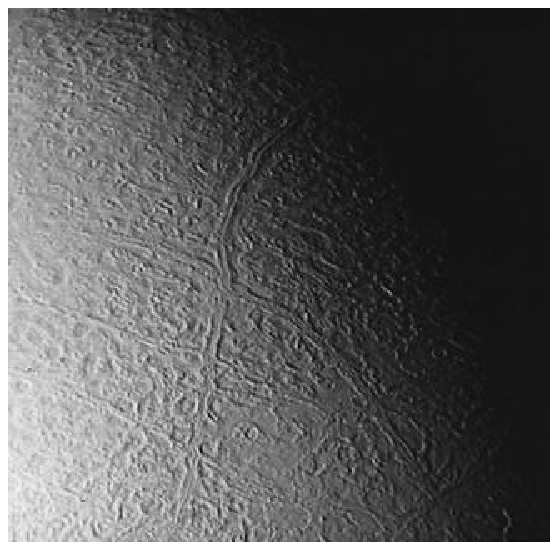} \\
\end{tabular}
\caption{The surface of Triton is merely sparsely cratered, thereby indicating past intense geologic activity \citep{Smith-1989}.
(Top left) High-resolution surface mosaic of Triton's anti-Neptunian hemisphere taken by Voyager 2 imagery; (top right) dark spots in the southern polar terrain caused by episodic geyser activity (spatial resolution 900m/pixel); (bottom left) possible cryovolcanic flow features; (bottom right) "Cantaloupe terrain" imaged from a distance of 130,000 km.}
\label{fig:OSS_Triton}       
\end{figure}

Another major science goal is to unravel Triton's geological history based on imaging science observations with high spatial resolution. 
This is prevented to date because of the limited coverage ($\sim$ 40\%) and low spatial resolution of most images collected during the Voyager flyby. 
The few higher resolution images reveal a geologically young, complex surface unlike any other seen in the outer solar system \citep{Prockter-2006}. 
A number of Triton's surface features (Figure \ref{fig:OSS_Triton}) are supposed to be of cryovolcanic origin \citep{Croft-1995}. 
Evidence for recent cryovolcanic activity was found in Triton's southern polar region, where erupting plumes were observed by the Voyager spacecraft \citep{Soderblom-1990}. 
The plume activity might be solar-powered, driven by seasonal sublimation and storage of nitrogen under translucent ice, then pressurized, and eventually released \citep{Kirk-1990}. 
Triton's cold surface (38 K) is covered by various volatile ices (N$_2$, CH$_4$ , H$_2$O, CO, and CO$_2$) that support the satellite's tenuous atmosphere being subject to seasonal variations \citep{Grundy-2010}. 

Triton's complex seasonal cycle puts Triton in a key position for the understanding of surface-atmosphere interactions on small icy bodies like Pluto and Eris with similar surface volatile inventories \citep{Abernathy-2009, Merlin-2009}. 
Radio science observations revealed a significant ionosphere with a well-defined peak at $\sim$350~km altitude \citep{Tyler-1989}. 
The distance and the geometry of the Triton closest approach precluded in situ observations of either the ionosphere or its interaction with Neptune's magnetosphere.

\subsection{Neptune's rings and inner satellites}\label{subsection:rings}

The OSS science payload and flyby trajectory offer a unique opportunity to increase our understanding of the Neptunian ring system and its retinue of small inner satellites. 
Ring-moon systems were once perceived as stable and unchanging for time scales of at least 10$^6$ to 10$^8$~years. 
New, higher-quality data from Cassini, Hubble and ground-based telescopes are painting a different picture, in which the systems evolve over years to decades. Cassini images even show changes in the F ring on a scale of hours to days \citep{Murray-2008}. 
High-resolution imaging and stellar (UVS) occultations can provide precise locations and shapes of known rings, detect new rings, and help to resolve the mechanisms behind the changes seen since the Voyager flyby. 
The knowledge gained by studying this system of rings and satellites will be applicable to ring systems and planetary disks that occur elsewhere in the universe.

\citet{Nicholson-1995} extrapolated the arcs' motion backward from the Voyager epoch and showed that all prior occultations were compatible with them, implying longevity of $\geq 5$~years \citep{Burns-2006}.
Without a confinement mechanism, arcs should disperse in a matter of weeks.
The nearby moon Galatea was quickly recognized as playing a major role in confining the arcs via a corotation resonance \citep{Goldreich-1986, Porco-1991}. 
However, the \citet{Goldreich-1986} corotation-sites are in fact unstable when solar radiation forces on dust are taken into account \citep{Foryta-1996}.
Thus the dust in the arcs must be replenished by macroscopic source particles. 
A good coverage of phase angles in the OSS flyby, extending to high phases greater than $155^{\circ}$, will be the key to infer the size distribution from visual and near-IR observations of the rings.
The dust detector can determine directly the composition of micron sized grains in the extended Neptunian dust disk
(detected by Voyager, \cite{Gurnett-1991}) from in situ measurements.
Since the dust particles are released
from larger bodies in the system, which ultimately are also the parent bodies of the rings, these measurements uniquely
constrain the composition of the whole Neptunian ring moon system.

\subsection{Kuiper Belt objects}\label{subsection:kuiper}

Our understanding of the history of our solar system has been revolutionized, largely because of the discovery of Kuiper Belt Objects (KBOs, \citet{Jewitt-1992}). 
Well over a thousand KBOs have since been discovered (e.g. \citet{Barucci-2008}), and they exhibit remarkable diversity in their properties. 
Geometric albedos range from a few percent to nearly 100\%, and appear to be correlated with diameter, as well as (possibly) visible color and perihelion distance \citep{Stansberry-2008}; cold-classical KBOs appear to be red and have high albedo \citep{Doressoundiram-2008, Brucker-2009}. 
Visible colours span the range from slightly blue to the reddest objects in the solar system (like for Pholus). 
Visible and near-IR spectra run the gamut from profoundly bland to exhibiting absorptions (or emission peaks) due to organics, water, nitrogen, methane and other hydrocarbon ices, and silicates (e.g. \citet{Barucci-2008b}). 
Perhaps the most remarkable character of KBOs is the abundance of binary and multiple systems: among the cold-classical KBOs, 30\% or more are probably binaries \citep{Noll-2008}.

The current number of KBOs, the dynamical structure of their orbits \citep{Kavelaars-2008}, and the diverse physical characteristics of the individual objects has spurred intense efforts at modelling the formation and evolution of the KBO population.
Beginning with the \citet{Malhotra-1993} explanation for Pluto's orbit, these studies have led to
a picture in which Saturn crossed through the 2:1 resonance with Jupiter, exciting the orbital eccentricities of Uranus and Neptune, which (as a result) interacted strongly with the primordial Kuiper Belt, suffering significant outward orbital migration.

In many areas, the KBO science objectives are similar to the Triton science objectives. 
Our overall objective is to guarantee that we obtain the necessary data for detailed comparison of Triton and the OSS KBO to each other, and to Pluto and the New Horizons KBO. Together, these observations will begin to provide significant insights into the diverse KBO population.
Because target selection likely will not have been finalized before detailed instrument definition work must be completed, the instruments need to have broad capabilities appropriate for the exploration of primitive bodies in the outer solar system. 
Luckily, Triton serves as an excellent proxy to a KBO target, and we do possess some detailed knowledge about a few of the larger KBOs (e.g. spectral features, albedos).
Starting with that knowledge, we developed measurement objectives tailored for Triton and the largest KBOs, and then enhanced those to include conditions and objectives appropriate for a range of KBOs (e.g. lower albedo surfaces, more tenuous atmosphere).
	
\section{Proposed payload}\label{section:payload}

For planetary objectives, the requirements on the instruments are equivalent to those of the New Horizon mission, with high technological readiness level (TRL) \citep{Weaver-2008}.
The technological progress and the new instruments with respect to Voyager 2 will ensure the scientific return of the mission. 
The performance requirements are more severe for fundamental physics objectives, with a modest impact on the spacecraft design.

The payload mass deduced from launcher capability is about 48 kg, less than the sum of all proposed instruments (see in Table \ref{tab:Inst}). 
So, the further step of mission analysis shall define a core payload coherent with the spacecraft design.

\begin{table}[h]
\caption{OSS strawman instrument payload}
\label{tab:Inst}
\begin{tabular}{|c|c|c|c|}
\hline
Instrument & Acronym & Mass (kg) & Power (W) \\
 \hline
Accelerometer & ACC      & 3.5  & 3.0 \\
\hline
 Radio-Science & RSI       & 3.0  & 40.0 \\
\hline
Ultra-Stable Oscillator & USO       & 1.5  & 5.5 \\
\hline
Very Large Base Line Interferometer & VLBI      & -    & 1.0 \\
\hline
Ultraviolet imaging spectrometer & UVS      & 5.0    & 12.0 \\
\hline
Near infrared imager & NIR       & 10.1 & 7.5 \\
\cline{1-2}
Wide angle camera & WAC       &      & \\			
\hline
High resolution Narrow Angle Camera &  NAC       & 9.8  & 14.0  \\
\hline
Radio and Plasma Wave & RPW       & 4.7  & 5.9 \\
\hline
Magnetometer & MAG       & 3.3  & 3.0  \\
\hline
Thermal imager & TMI       & 7.0    & 20.0 \\
\hline
Dust Particle Detector & DPD       & 3.5  & 9.0  \\
\hline
Laser-Science & LSI 2-ways & 25.0   & 80.0  \\
\cline{2-4}
 & LSI 1-way  & 14.5 & 21.5  \\
\hline
\end{tabular}
\end{table}

\subsection{ACC - Accelerometer GAP}\label{subsection:ACC}

The Gravity Advanced Package (GAP) \citep{lenoir2011electrostaticASR} complements the navigation instruments -- LSI, RSI and/or VLBI -- by measuring without bias the non-gravitational acceleration of the spacecraft. It provides an additional observable and allows removing, during the orbit restitution process, the effect of the non-gravitational forces on the trajectory, enhancing deep space gravitation tests as well as gravity field recovery \citep{lenoir2010odyssey2IACproceedings}.

GAP is composed of an electrostatic accelerometer MicroSTAR, based on Onera expertise in the field of accelerometry and gravimetry with CHAMP, GRACE, and GOCE missions \citep{Touboul-1999}, and a Bias Rejection System. Ready-to-fly technology is used with original improvements aimed at reducing power consumption, size and mass.

\begin{figure}
 \begin{tabular}{cc}
  \includegraphics[width = 0.35 \linewidth]{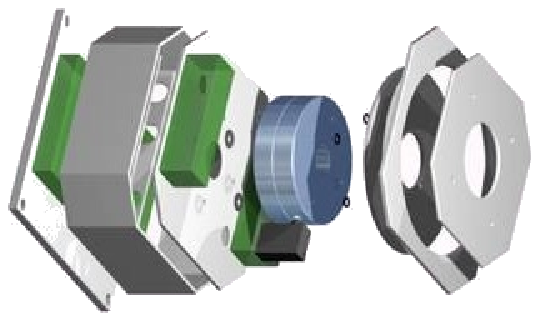} &
  \includegraphics[width = 0.6 \linewidth]{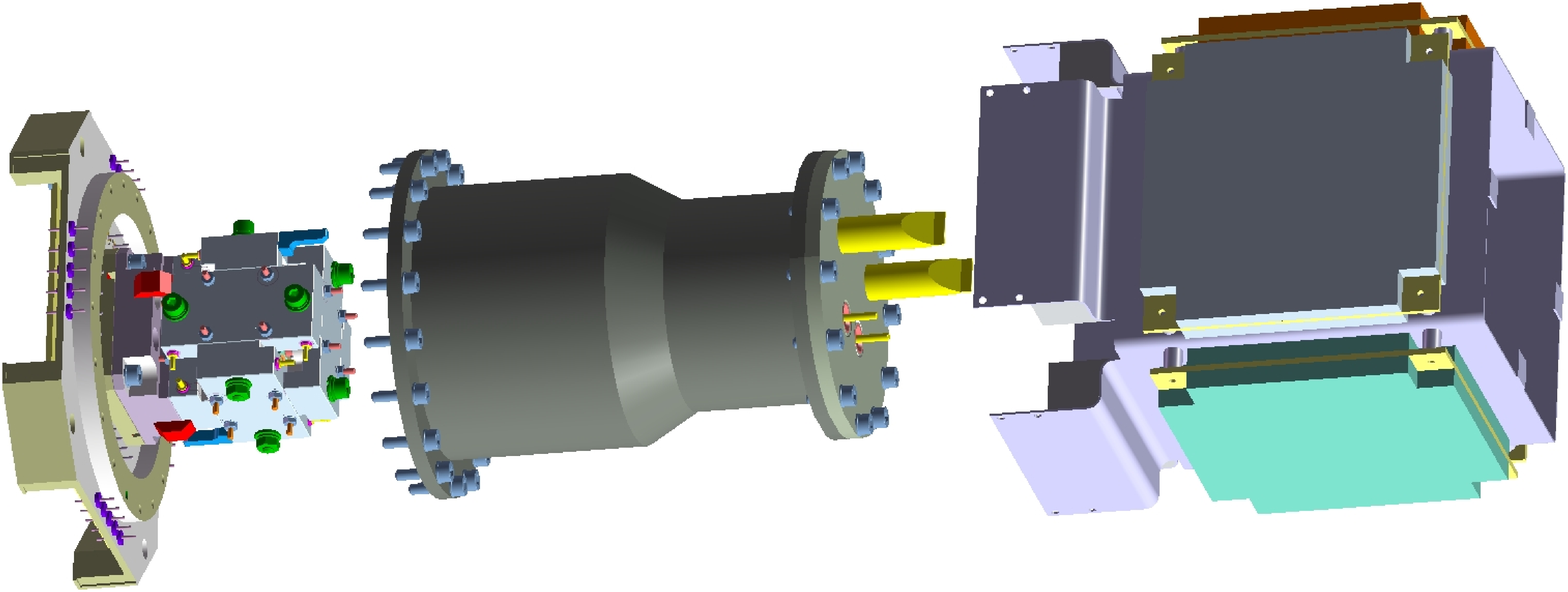} \\
 \end{tabular}
 \caption{Exploded view of the Gravity Advanced Package: Bias Rejection System (left) and MicroSTAR accelerometer (right)}
 \label{fig:ACC}
\end{figure}

MicroSTAR is a three axes accelerometer using the electrostatic levitation of a proof-mass with a measurement range of $1.8\times10^{-4}$ m~s$^{-2}$. The proof-mass is controlled by electrostatic forces and torques generated by six servo loops. Measurements of these forces and torques provide the six outputs of the accelerometer. The mechanical core of the accelerometer is fixed on a sole plate and enclosed in a hermetic housing in order to maintain a sufficient vacuum condition around the proof-mass (cf. fig.~\ref{fig:ACC}). The electronic boards are implemented around the housing with low consumption analog functions.
The Bias Rejection System \citep{selig2011technologyIACproceedings}, which is a rotating platform, is used to remove the bias of MicroSTAR along two perpendicular axis in the orbit plane. It is composed of an angular actuator working in a closed loop with a high-precision encoder and the overall angular accuracy is equal to $10^{-5}$~rad.

In order to remove properly the bias of MicroSTAR, the Bias Rejection System rotates the accelerometer following a carefully designed periodical pattern \citep{lenoir2011measuringSF2Aproceedings}. In terms of measurement noise, this operation selects the noise of MicroSTAR at approximately the modulation frequency. After post-processing and for a modulation period of 10 min, it allows making absolute measurements with a white noise whose Power Spectrum Density (PSD) level is $10^{-10}$~m~s$^{-2}$~Hz$^{-1/2}$ (same level as GRACE accelerometer) with a cut-off frequency equal to $8.3\times10^{-4}$~Hz \citep{lenoir2011unbiased}. This corresponds, for an integration time of 3 hours, to a precision of 1~pm/s$^{2}$ and an exact measurement accuracy. The methodology used to derive this precision level from the PSD of MicroSTAR has been validated experimentally. When taking into account the integration of the instrument in the spacecraft, and in particular the alignment accuracy ($\leq$ 1 mrad), the positioning accuracy and the spacecraft self-gravity, a global precision of 10~pm/s$^{2}$ is expected.

\subsection{LSI - Laser Science Instrument}\label{subsection:LSI}

For the verification of the Eddington's parameter $\gamma$ at $10^{-7}$, it is proposed to primarily use laser ranging/Doppler measurements instead of radio-science observations in order to achieve an accuracy of the Doppler measurement of $10^{-16}$ over the test duration. Two solutions are proposed, the first one based on a one-way pulsed laser concept (TIPO) and the second one more ambitiously based on a two-way coherent continuous laser concept (DOLL). 

\subsubsection{One way Laser - TIPO}

The TIPO experiment (T\'el\'em\'etrie Inter Plan\'etaire Optique) proposed by the OCA team is a one-way laser ranging project derived from satellite and lunar laser ranging (SLR/LLR) and optical time transfer T2L2 \citep{Fridelance-1997}. The TIPO principle is based on the emission of laser pulses from an Earth based station towards the spacecraft. These pulses are timed in the respective timescales at departure on Earth and upon arrival on the spacecraft. The propagation time and the respective distance between Earth and spacecraft are derived from the difference of the dates of departure and arrival. This one-way laser ranging permits distance measurements on a solar system scale (up to 30~AU) as the one-way link budget varies only with the square of the distance, contrary to the power of four for usual laser telemetry.

\begin{figure}
\begin{tabular}{cc}
  \includegraphics[width = 0.45 \linewidth]{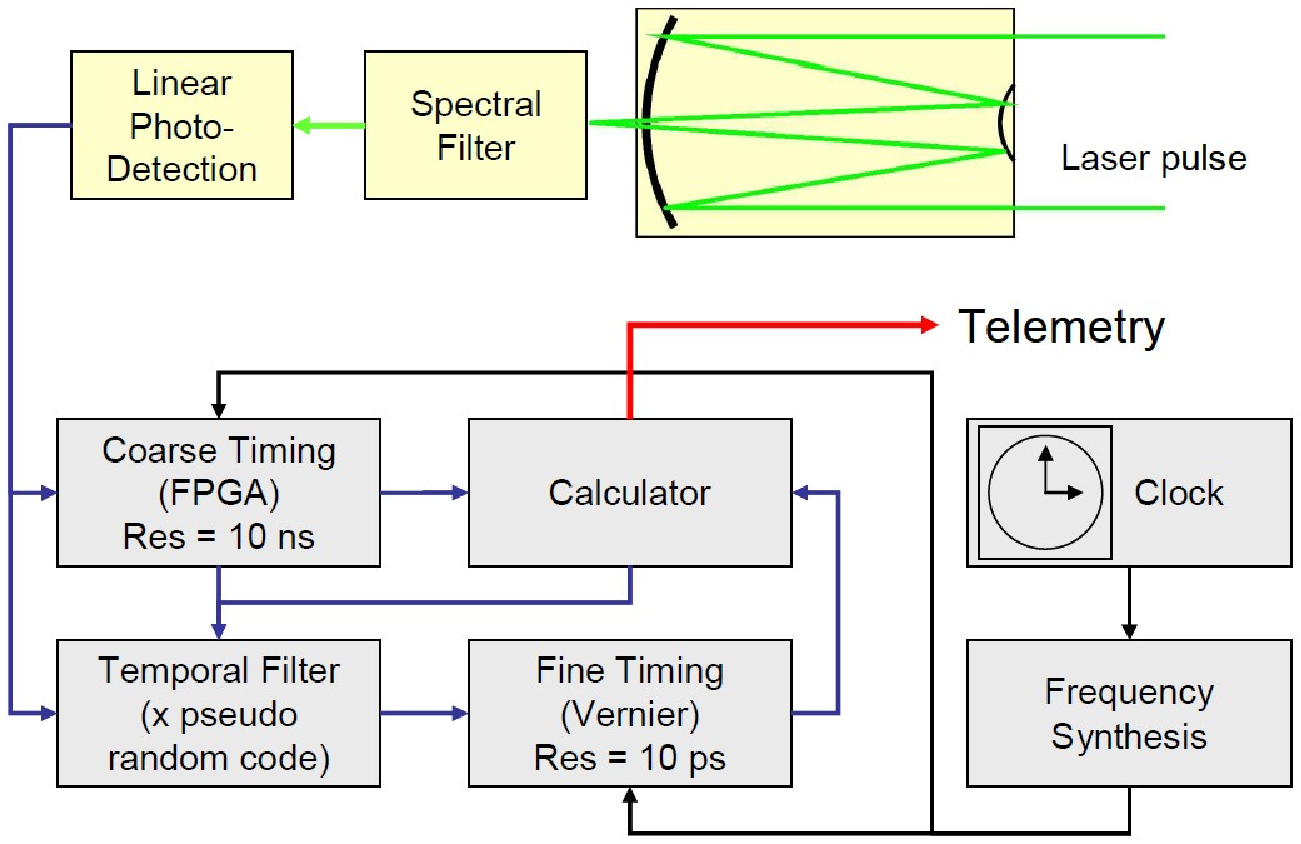} & 
  \includegraphics[width = 0.45 \linewidth]{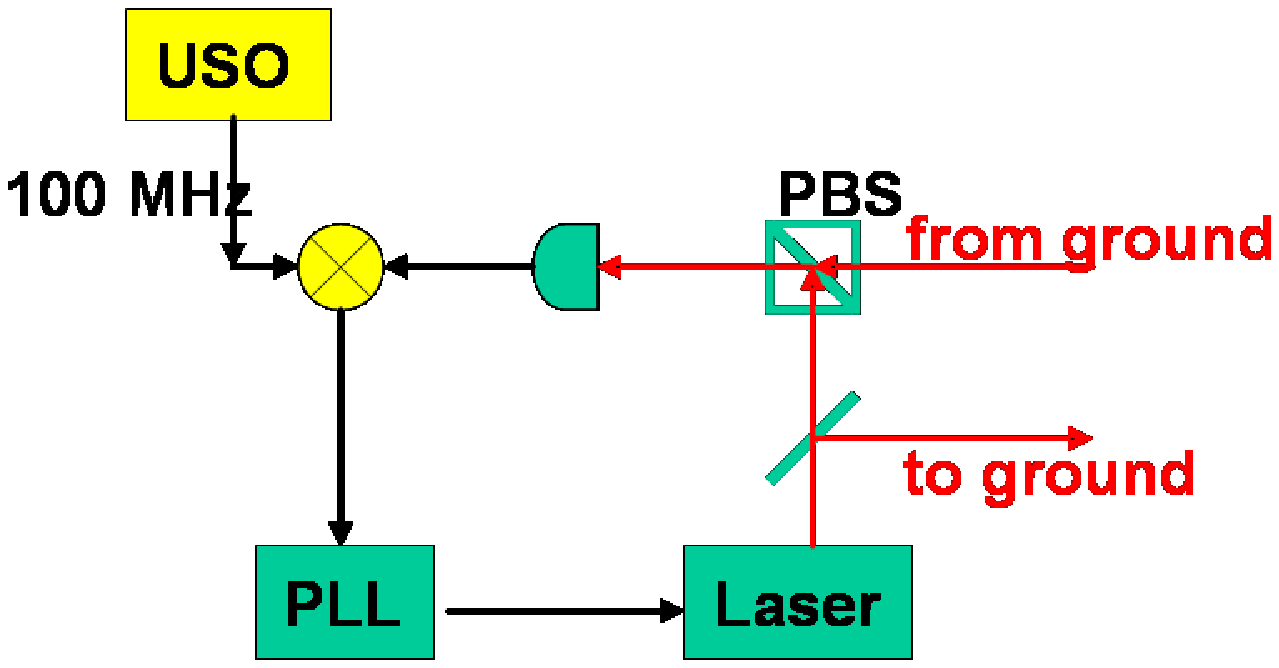} \\
\end{tabular}
\caption{TIPO (left) and DOLL (right) space segment synopsis}
\label{fig:LSI-1}       
\end{figure}

The ground segment is a high-fidelity laser ranging station, working solely as an emitter. 
An example of a suitable ground station is the rejuvenated Ex-LaserLune station "MeO" of OCA in Grasse, France, but there is half a dozen suitable laser ranging stations worldwide that fully (or with minor enhancement) comply with the requirements (laser power, telescope diameter and pointing performance). 
The space equipment consists of an optical subsystem (small telescope and detection device), an electronic subsystem (event timer, detection and control electronics) and a clock (USO - Ultra Stable Oscillator), as illustrated in Figure \ref{fig:LSI-1} (left panel). The optical subsystem comprises a rather compact telescope in order to collect the incoming laser pulses, a spectral filter for noise reduction and a linear photon detection system with picosecond timing characteristics. Considering maximum distance of 30~AU, a 20~cm aperture telescope and an acquisition phase of 1000~s, the signal to noise ratio may be raised to a satisfying signal confidence level of 1 to 10.

Considering an onboard clock with an Allan variance better than $10^{-15}$~@~100~000~s (cold atomic clock HORACE designed by SYRTE \citep{Esnault-2011} or the mercury ion clock designed by JPL \citep{Prestage-2007}), TIPO allows differential distance measurements accurate to a millimetre, leading to equivalent Doppler measurement in the range of $10^{-16}$ over one day. 

\subsubsection{Two way coherent laser - DOLL}

The DOLL optical link concept for OSS (Deep space Optical Laser Link) is the optical equivalent of the radio link, with an on-board laser transponder (Figure \ref{fig:LSI-1}, right panel) and ground terminals at already operating satellite/lunar laser ranging stations. 
The optical link is based on the coherent optical link for navigation and timing initially envisaged for SAGAS \citep{Wolf-2009}, but is making use of existing flight hardware developed by TESAT GmbH for the Laser Communication Terminal (LCT) presently flying onboard the German TerraSAR-X and the US NFIRE satellites \citep{Gregory-2010}.

There are 3 main issues for achieving the scientific objectives : the number of photons arriving to the telescope, due to the distance; the stray light from the sun; the atmosphere turbulence. For solving these issues, the baseline version OSS-DOLL features in particular:
\begin{itemize}
	\item Continuous wave laser operation in both directions (two-way system) at $\lambda$=1064.5~nm;
	\item Heterodyne onboard laser transponder (minor modification of the present homodyne LCT transponder);
	\item Large stray light rejection from heterodyne detection and due to a controlled frequency offset (100 MHz) between incoming and outgoing laser signals, using a space qualified USO (ACES/PHARAO Quartz oscillator);
	\item Adaptive optics methods to ensure phase coherence over the complete aperture of the ground telescope
\end{itemize}

Thanks to narrow band pass interference filters, the 5 kHz heterodyne detection filter and the reduction by the ratio of "laser spot size" (diffraction limited to about $5 \times 10^{-6}$ rad) to "Sun spot size" (about $5 \times 10^{-3}$ rad at 2 AU), only $5\times 10^{-16}$~W of the stray light will arrive to the detector, well below the signal at $10^{-14}$~W (with 1 W emitting laser and 20 cm diameter telescope on board, 1.5 m telescope on ground and assuming a beam pointing efficiency of 0.8, a transmission of the Earth atmosphere of 0.9 and of the instrument of 0.1 and a distance of 2  to 3 AU). 
Moreover, stray light from the outgoing signal into the detection channel is mitigated by using orthogonal polarizations, and by offsetting the outgoing frequency with respect to the incoming one.

The phase coherent detection for the DOLL concept requires also adaptive optics methods to ensure phase coherence over the complete aperture of the 1.5 m ground telescope \citep{Robert-2007}.
The wave-front sensor of the adaptive optics would use the incoming signal, which requires the development of a low power wave front sensor, operational in the presence of large background light. 
Such a sensor based on heterodyne interferometry (using a narrow electronic filter to reject stray light) is under development for different applications. 
Another advantage of such a detector is the possibility of using its output directly for the phase measurement, i.e. there is no need to "share" photons between the adaptive optics and science channel.

The dominating noise sources are the effects due to the atmosphere (turbulence and slow index variations), with a resulting overall power spectral density of $10^{-27}$/Hz at high frequency ($> 10^{-3}$ Hz), the transition to white phase noise (f$^2$ slope) at $10^{-3}$~Hz, and back to white $10^{-27}$/Hz white frequency noise at low frequencies ($< 10^{-5}$ Hz) from the mm tropospheric model errors (the noise model is based on measurement of atmospheric turbulence using optical stellar interferometry \citep{Linfield-2001} and ground links \citep{Djerroud-2010}).  Below $3\times 10^{-5}$~Hz, the onboard accelerometer noise becomes dominant, but it plays only a marginal role in the measurement uncertainty of the Eddington's parameter $\gamma$ as most of the signal variation is over a few hours around conjunction i.e. most of the signal energy is concentrated at frequencies above $10^{-4}$ Hz.
For one conjunction at 2 AU, $\gamma$ is determined with an uncertainty of $1.2 \times 10^{-7}$, using 10 days of data before and after the occultation.

\subsection{RSI - Radio-Science, USO - Ultra-Stable Oscillator and VLBI - Very Large Baseline Interferometer}\label{subsection:RSI}

The Outer Solar System (OSS) mission has radio science objectives in two categories, gravitation and propagation.

The gravitational measurements are based on precision determination of perturbation to the spacecraft motion or orbit as a result of gravitational forces acting on it from planets, moon, rings, or other bodies in its environment as well as relativistic effects. These measurements are made with precision Doppler tracking, which are optimized at two wavelengths, X- and Ka-bands, in a two-way coherent mode. The dual links enable the calibration of the dispersive effects of the charged particles in the interplanetary plasma. X- and Ka-bands have been flown reliably in a coherent mode on at least two deep space missions (Cassini \citep{Kliore-2004} and Juno) and are planned for several more upcoming missions (Bepi-Colombo \citep{Iess-2009}). The two-way coherent mode enables the transponder(s) to take advantage of the superior stability of H-maser based clocks at the ground stations. 
The performance of the radio links can be expressed in Doppler noise in units of velocity or in terms of the dimensionless Allan deviation and should be at least $10^{-14}$ at an integration time of 1000~s in order to ensure an accuracy of 0.003~mm/s \citep{Asmar-2005}.

The propagation radio science experiments measurements are based on precision determination of perturbation in the phase/frequency and amplitude of the radio signals propagating from the spacecraft to Earth by intervening media under study, such as atmospheres and ionospheres of the planets, moons, or other bodies. These measurements are made in the one-way mode utilizing a highly stable reference to generate the signal on-board the spacecraft at two or more links. The dual/multiple one-way links enable the isolation of dispersive material under study such as planetary ionospheres. Numerous deep space missions (Voyager, Galileo, Mars Global Surveyor, Cassini, GRACE, Rosetta, Venus Express) have successfully utilized this technique for occultation experiments by flying Ultra-Stable Oscillators (USO), which are quartz resonators in single or dual ovens for thermal stabilization.  The performance of the propagation links can be expressed in terms of the dimensionless Allan deviation of phase stability and the state of art is at least $10^{-13}$ at integration times between 10 and 1000~s \citep{Tyler-2008}.

Observations of the spacecraft using global VLBI arrays with 10 or more radio telescopes and maximum baselines of $\sim$10~000~km at X-band in a phase referencing mode using the natural extragalactic radio sources for phase calibration can provide positioning accuracy at a level of 0.1~nrad, or 150~m at a distance of 10~AU.
Regarding the on-board segment, no special requirements on top of the Radio Science and USO are implied. VLBI can be done on the regular transmission signals provided a power in the data carrier line and ranging tones (combined) will be a level of 1~W at 20~dBi TX antenna gain, ranging tones separation of $\sim$100~MHz and intrinsic frequency stability at a level of $10^{-12}$ at 100~s.
VLBI observations of the signals transmitted during the two-way link Radio Science experiments will improve the Doppler measurements by a factor of 1.4 due to the averaging of the propagation scintillations effects in Earth troposphere, ionosphere and interplanetary plasma.

\subsection{NAC - High resolution Narrow Angle Camera}\label{subsection:NAC}

The High Resolution Camera will conduct imaging science of multiple targets of interest in the Neptunian system.  It will be used to measure global cloud motions during the approach phase of the mission, map the fine-scale structure of its ring system, and produce high-resolution surface maps of Triton.
High resolution (colour) and panchromatic imagery of the Neptunian system is envisaged with 0.005~mrad spatial resolution. The optics will be designed as a reflective telescope with a large focal length and an imaging array sensor, which can be operated either in a pushbroom mode for high resolution panchromatic imaging of Triton during flyby (with flyby involved motion compensation for enabling long exposures) or in a conventional framing mode (for high resolution and color images of the Neptunian system and to obtain images to support spacecraft tracking). 

The instrument will consist essentially of three components, the optics, the sensor system, and a data processing unit. The optics will be a reflecting telescope involving a primary mirror and a secondary convex mirror. The focal length is 3~m.
The baseline sensor is a CMOS Star1000 1024 x 1024 array sensor (pixel size 15~$\mu$m), which is known to be radiation-resistant, and for which much heritage is available \citep{Hopkinson-2008,Defise-2004}. Optionally, a motorized filter wheel with 12 filter positions could be mounted in front of the entrance optics to allow for colour and multispectral observations of distant targets in the Neptunian system.

The instrument will operate during the tour through the Neptunian system as well the KBO observations. The pointing prediction shall be sufficiently accurate to successfully point the camera at distant targets.  The pointing shall be stable within the size of 1/3 image pixel during the typical exposure time of 0.5~s. During orbit, the camera shall maintain nadir-pointing.  The direction of the motion vector must be held throughout the orbit mission. The instrument operation schedule should allow for geometric and radiometric calibration, instrument alignment cross-calibration and performance tests.
The calibration is done by measurements of instrument alignment with the spacecraft coordinate system using star observations and radiometric calibration of the camera using star observations.

The performance characteristics of the instrument are:
\begin{itemize} 
\item Spectral range: 350 - 1050~nm;
\item Field of view: 0.293$^{\circ}$; 
\item Pixel field of view: 0.005~mrad
\end{itemize}

\subsection{MAG - Magnetometer}\label{subsection:MAG}

The magnetometer will measure the magnetic field vector in the bandwidth DC to 128~Hz. It is composed of at least one but preferably two tri-axial fluxgate sensors, which would be boom mounted in order to minimise magnetic interference, and a platform mounted electronics box.

\begin{figure}
  \includegraphics[width = 0.5 \linewidth]{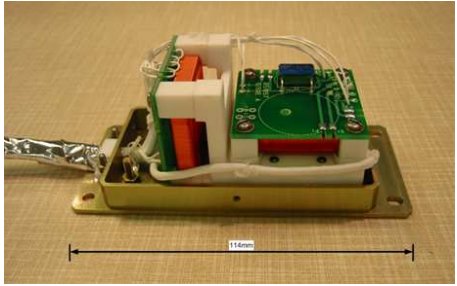}
\caption{Double Star fluxgate sensor}
\label{fig:MAG}       
\end{figure}

Two sensors are preferred to facilitate operation as a gradiometer in order to separate the very small target ambient field changes from any magnetic disturbance field due to the probe fields (see Ness et al (1971), Georgescu et al (2008)). 
The sensors would be ring core fluxgates, similar to that shown in Figure \ref{fig:MAG}. Fluxgate sensors have flown on many space plasma and planetary missions (e.g. Galileo, Cassini, Cluster, Rosetta, THEMIS) and can thus draw on considerable space heritage \citep{Acuna-2002, Glassmeier-2007, Balogh-2010}. 
Modern sensors feature very low noise ($\leq$ 10pT/$\surd Hz$) above 1~Hz, good offset stability ($\leq$ 0.05 nT/K) and scale factor drift ($\leq$ 40 ppm/K)) \citep{Carr-2005}.

The fluxgate is implemented within the standard feedback loop (housed on the electronics card) featuring bipolar driving of the soft magnetic ring core and second harmonic detection of the field proportional feedback voltage.  The sensor electronics would be a digital FPGA based design (direct heritage from Bepi-Colombo and THEMIS, \citet{Glassmeier-2010, Auster-2008}), or an ASIC which would require further specific development but offer a reduction in instrument power consumption. 

The magnetometer has no pointing or active alignment requirements; however accurate knowledge of the sensor orientation (to better than $0.1^{\circ}$) is required in order to accurately determine the field direction. 

The magnetometer would be calibrated on the ground prior to launch. In-flight calibration would utilize techniques developed on previous missions \citep{Gloag-2010, Leinweber-2008, Kepko-1996} using a combination of data taken during Earth fly-bys, passage through the solar wind and Fourier analysis of data acquired during spacecraft spin-modes. The calibration analysis will determine changes to parameters in the sensor calibration matrix and, via the gradiometer, determine and subtract any perturbing spacecraft field.

One issue with the fluxgate sensor is the cold environment at the outer planets in the event that no power resource is available for MAG sensor heating. 
It is predicted that the temperature in the environment of boom mounted sensor in Neptune orbit could be as low as 70~K. 
There are currently technology studies underway at Imperial College London into cold running of fluxgates as part of the ESA-JUICE mission instrument studies.

\subsection{RPW - Radio and Plasma Wave}\label{subsection:RPW}

The RPW experiment will provide remote and in situ measurements of radio and plasma wave phenomena in the frequency range from a fraction of a Hz up to about 20-40~MHz for electric fields and 100~kHz for magnetic fields. The scientific objectives of this experiment are (i) the determination of the Poynting vector and the full polarization state of electromagnetic waves for remote sensing of Neptunian (NKR) and heliospheric electromagnetic emissions, (ii) a high frequency coverage up to 20-40~MHz to sample a significant portion of the spectrum of NEDs (Neptunian Electrostatic Discharge), whose low frequency cut-off would provide a remote sensing tool of the sub-spacecraft electron peak ionospheric density and (iii) the detailed study of local wave phenomena and the identification of characteristic frequencies of the local plasma. 

RPW will include three 5-m long electric orthogonal monopole antennas and three magnetic orthogonal search coils. After pre-amplification, the sensors outputs are processed by a low-frequency receiver from 0 to 20~kHz including a waveform sampler, and a high-frequency receiver from 10~kHz to 20~MHz able to measure auto- and cross-correlations of signals simultaneously sensed on two channels, stored in an electronic box. The experiment is powered by DC/DC converter and controlled by a central microprocessor (Digital Processing Unit) which will be used in flight to configure the operation modes (integration time, spectral bandwidth, spectral resolution, number of selected sensors for simultaneous measurements) in order to maximize the science return with respect to available bit rate and power. 

Such an experiment has a high maturity with strong space european heritage (Cassini/RPWS \citep{Gurnett-2004}, STEREO/Waves \citep{Bougeret-2008} and ongoing developments for Solar Orbiter/RPWI, BepiColombo MMO/RPW/Sorbet and JUICE/RPWI).

The RPW instrument can operate in many modes with various time and frequency sampling characteristics, that can be remotely reconfigured at any time to adapt to new scientific objectives. There will be onboard processing: either for onboard key parameter computation, for event triggering, or for lossless compression. Based on Cassini/RPWS, where the telemetry rate typically reaches a few kbps, the duty cycle will be adapted to match the available telemetry rate. 

RPW electric and magnetic sensors are sensitive to electric and magnetic interferences. While the radio antenna will be fixed directly on the spacecraft body, magnetic search coils need to be placed on a boom, that can be that of the magnetometer.
Several calibration procedures will be conducted: ground- based calibrations of the receiving chain (noise level, gain and phase), ground based calibration of sensors (effective length and direction of sensors through rheometry or wire-grid modeling) and in-flight calibration (with internal or external known sources). 

\subsection{NIR - Near infrared imager and WAC - wide angle camera: Norton}\label{subsection:NIR}

Norton will be used to image Neptune during the closest approach to capture the detailed global view of atmospheric cloud structure while simultaneously resolving the small eddy and aerosol structures using multiple filters.
This instrument is heavily based on the Ralph instrument of the New Horizons mission \citep{Reuter-2008}, and inherits much of the architecture and technology of that instrument system with modifications to the detector and filter systems. 
The comparable performance demands on the New Horizons spacecraft to those of the OSS mission result in a convergent evolution with the near-IR mapping spectrometer sharing optics with the wide-angle imaging camera.  
This results in a considerable mass savings, as the telescopes are a significant mass of an imaging instrument, especially for the low illumination and longer flyby distances in the outer solar system.

Norton shares the same optical design as Ralph, a single highly baffled 75~mm aperture in front of an f/8.7 visible/near-IR off-axis three-mirror telescope. 
This design provides good thermal and alignment stability, and adequate light throughput. 
Stray light is controlled not only via baffling, but also a Lyot stop at the exit pupil and further baffling at an intermediate focus. 
At the end of the telescope a dichroic beamsplitter delivers the light from wavelengths longer than 1.1~$\mu$m to the near-IR focal plane while shorter wavelengths are sent to the visible focal plane.

The visible focal plane is almost identical to the Multi-spectral Visible Imaging Camera (MVIC) sub-instrument from New Horizon's Ralph, with 9 independent 5024x32 CCD arrays operated in time delay integration (TDI) mode. 
Two of those arrays are used to produce panchromatic imagery (400-975~nm) while the other 7 are combined with filters to produce images in discrete bandpasses, blue (400-500~nm), green (500-600~nm), red (600-700~nm), near-IR (700-975~nm) and a narrow band methane (860-910~nm) and two nearby continuum channels (810-860~nm and 910-960~nm).
The angular resolution of each pixel in the visible focal plane is 20x20~$\mu$radian$^2$, and the static field of view (FOV) of the TDI array is 5.7$^{\circ}\times $0.037$^{\circ}$, the same as Ralph's MVIC. 
For targets like Neptune or Triton, Norton's visible focal plane will yield images with a signal to noise higher than 48 for all bandpasses. 

The near-IR focal plane is also quite similar to the Linear Etalon Imaging Spectral Array (LEISA) sub-instrument from New Horizon's Ralph, but with somewhat more extensive modifications. 
In particular, while the Ralph's LEISA instrument was sensitive from 1.25-2.5~$\mu$m, Norton's near-infrared focal plane will cover 1.25-4.5~$\mu$m. 
Additionally, technology now allows a much larger HgCdTe detector array (2048x2048~H2RG) to be used. 
The near-IR focal plane uses a linear variable filter superimposed on top of the detector to limit different columns of the detector array to be sensitive to different wavelengths. 
As the instrument is scanned past a scene, a particular region of the scene illuminates pixels sensitive to different wavelengths, thus building up a spectral image cube of the whole scene. 
Norton's near-IR spectral resolution will be R=600 throughout its bandpass, with a spatial resolution of 50x50~$\mu$radian$^2$, and a FOV width of 5.87$^{\circ}$. 
The SNR of Norton's near-IR focal plane for a target such as Neptune will yield a measurement of the reflectivity with a signal to noise ratio of about 30 throughout the bandpass.

\subsection{UVS - Ultraviolet imaging spectrometer: UVIS}\label{subsection:UVS}

The instrument measures photons in the 55-155~nm range with a spectral resolution of 0.54~nm (fwhm). The spatial resolution is 0.05$^{\circ}$, with a temporal resolution of 1s. The sensitivity is 0.47~count/s/cm$^2$ for extended source. The instrument can measure emissions from the atmosphere of Neptune and Triton, either nadir observations or airglow emissions of the upper atmosphere. Vertical sounding of major species can be obtained by stellar occultation.

\begin{figure}
	\includegraphics[width = 0.5 \linewidth]{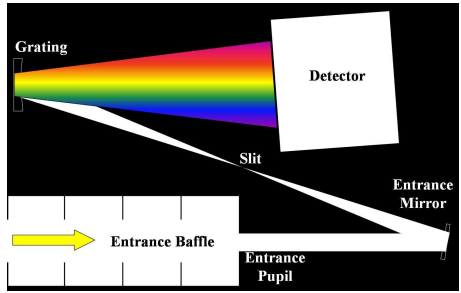}
	\caption{Optical concept of the UV spectrometer}
	\label{fig:UVS}       
\end{figure}

The instrument is a grating spectrometer, composed of a collecting part and a detector part. The collecting part is composed of an entrance baffle, a primary off-axis mirror and a slit. The detector part is composed of a grating and an intensified detector. The intensifier uses a CsI photocathode and a micro-channel plate in front of cross-delay resistive anode detector. The spacecraft interface is handled by a local DPU. The optical concept is shown in Figure \ref{fig:UVS}. 

Stellar occultations are performed by pointing the platform in a chosen direction and staying in inertial pointing for a few minutes. Other observations can be done either in inertial mode or in nadir pointing mode.
The required pointing accuracy is half the slit width (0.05$^{\circ}$) with a stability of 0.05$^{\circ}$ during the integration time (1~s). In inertial mode, the platform should be stable within one slit width (360~arc~sec) over a few minutes (for occultations).

A first calibration is performed on the ground. Evolution of the instrument sensitivity is measured in-flight by looking at stars. Some manoeuvres, such as rolls, are required to perform in-flight checks on straylight levels and polarization characterization.
The CsI photocathode must be kept in vacuum. The detector unit has a window which is opened once after launch. For ground activities, the window can be opened when in a vacuum tank. When the window is closed, the detector is pumped on a regular basis to maintain a sufficient vacuum level. 

The instrument has heritage from various existing or in-development instruments: PHEBUS on BepiColombo, SPICAM-UV channel on Mars-Express \citep{Bertaux-2000} and SPICAV-UV on Venus-Express \citep{Bertaux-2007}.

\subsection{TMI - Thermal imager OPTIS}\label{subsection:TMI}

OPTIS (Outer Planet Thermal Imager Spectrometer) is a thermal infrared imaging spectrometer with an integrated radiometer. The scientific goal of OPTIS is to provide detailed information about the mineralogical composition of an solid surfaces in the outer solar system by measuring the spectral emittance in the spectral range from 7-12~$\mu$m with a high spatial and spectral resolution. Furthermore OPTIS will obtain radiometric measurements in the spectral range from 7-40~$\mu$m to study the thermo-physical properties.

OPTIS builds on the heritage of MERTIS (Mercury Radiometer and Thermal Infrared Spectrometer) instrument for the ESA BepiColombo mission to Mercury \citep{Hiesinger-2010}. 
OPTIS uses a cooled MCT array detector to maximize the signal to noise ratio for the much colder surface of Triton and KBO. 
OPTIS has a FOV of 11.7$^{\circ}$ giving a much larger coverage of the surface within one orbit. 

The spectrometer of OPTIS is based on the Offner design with a micro-machined grating. In combination with the MCT detector OPTIS will cover the spectral range from 7-12~$\mu$m with a spectral resolution of 200~nm and a high spatial resolution. The radiometer is highly miniaturized and integrated in the slit plane of the spectrometer. The approach of combining a spectrometer and a radiometer with the same entrance optics provides synergies benefiting the scientific analysis. The fact that the surface temperature can be obtained independently from the spectral measurements allows removing ambiguities in the retrieval of emissivity values. The radiometer will further map thermal physical properties like thermal inertia, texture and grain size.

\begin{figure}
	\includegraphics[width = 0.7 \linewidth]{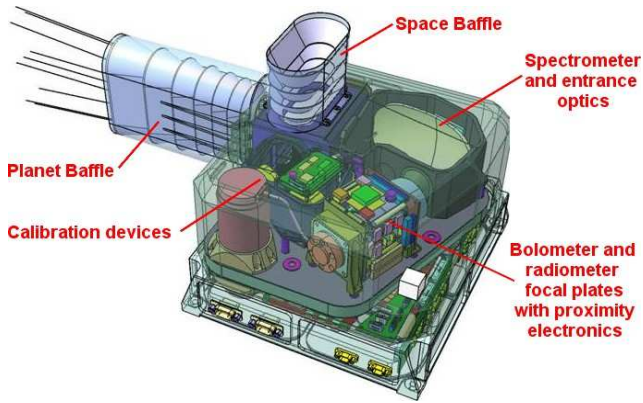}
	\caption{Concept views of OPTIS (view inside the instrument, thermal interfaces not shown)}
	\label{fig:TMI}       
\end{figure}

The instrument design is driven by a strong need for miniaturisation and a modular combination of the functional units at the same time. 
The sensor head structure (Figure \ref{fig:TMI}) contains the entrance and spectrometer optics, the bolometer and radiometer focal plates, its proximity electronics, the calibration devices (shutter and 300~K black body) and provides thermal interfaces to the spacecraft to achieve required high thermal stability. 
At its optical entrance a pointing device is located which directs the incoming infrared beam from 4 different targets, 3 for in-flight calibration purposes and the planet view itself: planetary body view, the look into deep space (space view), the image of the 300~K black body and the image of a spectral calibration target.

OPTIS will typically operate in a mapping mode. In this mode the instrument will alternate between the three calibration views and the planet view with a defined duty cycle. Spectral and radiometric data are obtained simultaneously. Binning in spectral as well as spatial direction can be performed in the instrument by software mode to optimize the signal to noise ratio based on the temperature of the target.

\subsection{DPD - Dust Particle Detector}\label{subsection:DPD}

The in-situ dust sensor is based upon impact ionization and measures size, speed, direction, and chemical composition of individual dust grains. An interplanetary spacecraft to Neptune provides the unique possibility to link inner solar system dust (processed) with outer solar system dust (Kuiper belt particles) properties. Furthermore, Neptune's variable dusty ring system is of particular interest and its properties like extension, density, variability, and composition shall be studied and compared to the Jovian and Saturnian ring systems, which have been studied by similar dust detectors. At a close flyby the detector encounters material lifted up from Triton's surface by micro meteoroid bombardment. It thus offers the unique opportunity of compositional in situ studies of Triton surface. The novel dust telescope, a successor of the instruments flown aboard Stardust \citep{Kissel-2003}, Galileo \citep{Gruen-1992a, Gruen-1992b} and Cassini \citep{Srama-2004} can operate during cruise and encounter phases.

\begin{figure}
  \includegraphics[width = 0.5\linewidth]{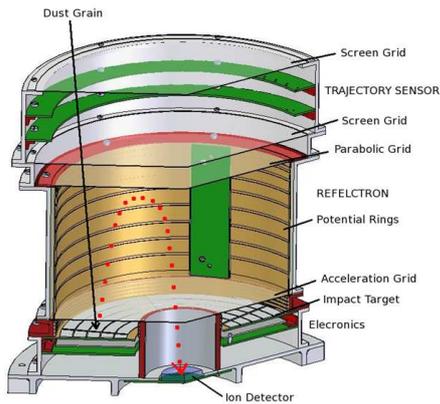}
\caption{Dust Particle Detector}
\label{fig:DPD}       
\end{figure}

The instrument measures low dust fluxes (interplanetary micrometeoroid background) as well as high impact rates e.g., when crossing ring segments. Therefore the dust instrument package operates in two modes: the cruise mode and the encounter mode. The cruise mode is optimised to measure low dust fluxes down to $\sim 10^{-4}$~m$^2$/s, small particle sizes ($\geq$~10-15~g), high impact speeds ($\geq$~2 - 50~km/s), grain primary charge ($\geq$~0.5~fC) and the particle composition of individual dust grains. The encounter mode provides reliable information about the dust environment with its high fluxes.  In order to achieve this wide range of measurement parameters, the instrument packages combines impact ionisation, TOF spectrometry and charge induction as detection methods. A basis of this multi-coincidence detector is a small dust telescope using ring shaped target segments. 
Two planes of charge sensitive wires called Trajectory Sensor (Figure \ref{fig:DPD}) at the instrument aperture measures grain primary charges whereas the impact is triggered by the grain hitting a target segment.
The integrated Sensor package has a field-of-view of $\pm 40^{\circ}$. 

\section{Mission profile and spacecraft design}\label{section:mission}

The capability to embark the instrument suite defined in the previous chapter is highly dependent on the chosen orbit and launcher. The following section presents a preliminary analysis of the mission profile showing the capability to deliver a 500 kg class probe. Then the design of the spacecraft is briefly described with the main characteristics required either by the science goals or by the mission profile.

\subsection{Mission profile}\label{subsection:orbit}

The selected orbit shall meet the following criteria:
\begin{enumerate}
	\item Long ballistic periods to follow geodesic arcs to as large a heliocentric distance as possible for gravity measurements,
	\item Sun occultation for Eddington parameter measurement,
	\item Neptune and Triton flybys for planetary science measurements, then a flyby of a scientifically selected Kuiper Belt object,
	\item Transfer to Neptune in less than 13~years for a rapid science return,
	\item Mission $\Delta$V, and thus the onboard propellant quantity, as small as possible to increase the delivered mass,
	\item Low departure velocity to reduce launch cost.
\end{enumerate}

The first requirement is naturally achieved by Neptune objective, leading to heliocentric distance larger than 30 AU, with possible extension after Neptune flyby.

For the second requirement, as Neptune's orbit is not in the ecliptic plane, the Earth-Sun-spacecraft conjunctions can occur only early in the mission or when the spacecraft is near the ecliptic plane.
With direct transfers, two conjunctions will occur, at 2.1~AU (6 months after departure) and at 4.3~AU. 
An indirect transfer increases the number of solar conjunction during inner solar system trajectory.

The Neptune encounter provides access to a huge cone of trans-Neptunian space in order to achieve the third requirement.
The bending angle that can be achieved varies with the velocity of approach to Neptune and the trajectory's minimum distance from Neptune. Faster and farther decrease the angle; slower could increase it. 
Trajectory modeling shows that tens of known KBOs are accessible to OSS, and the flyby geometry can be tailored to not only achieve science goals within the Neptune system, but continue on to a scientifically-selected Kuiper Belt Object afterward \citep{Marley-2010}. 

\begin{figure}
\begin{tabular}{cc}
  \includegraphics[width = 0.45 \linewidth]{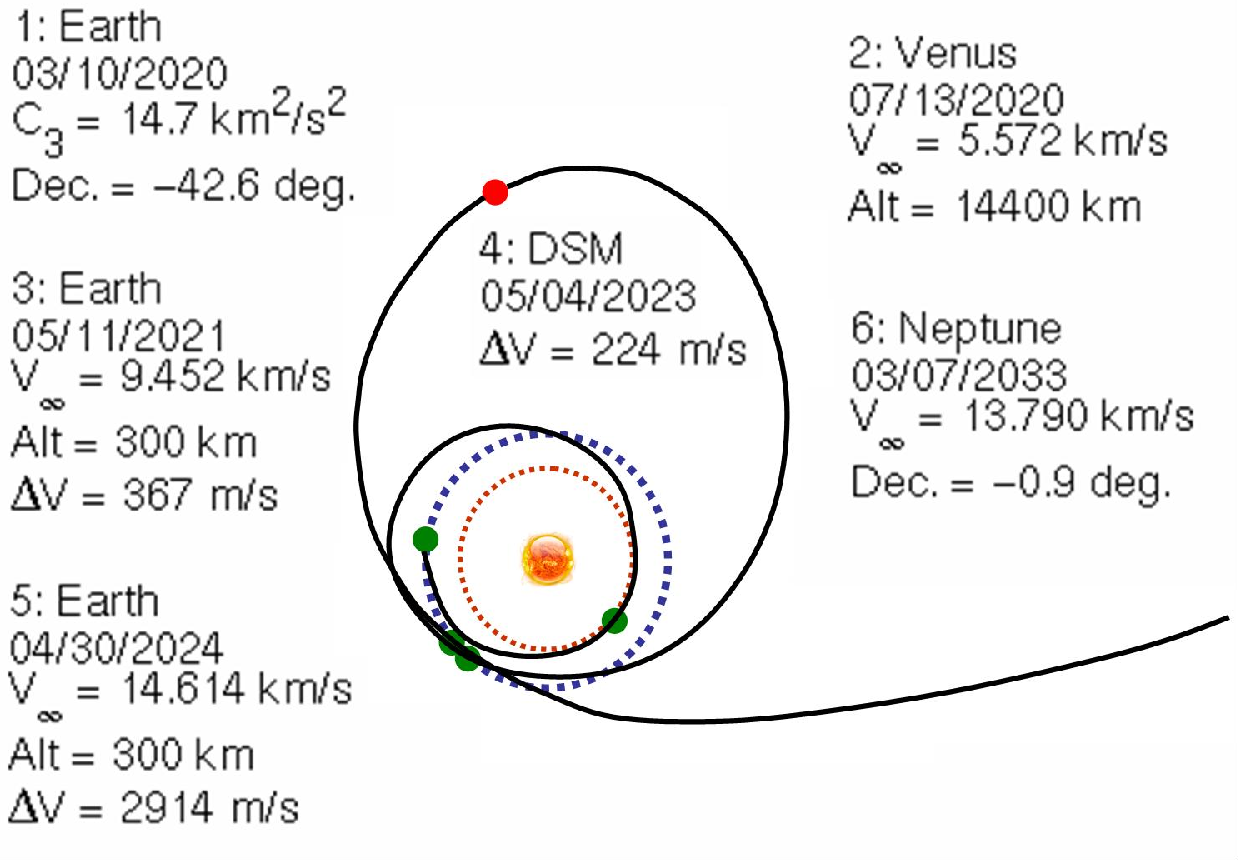} & 
  \includegraphics[width = 0.45 \linewidth]{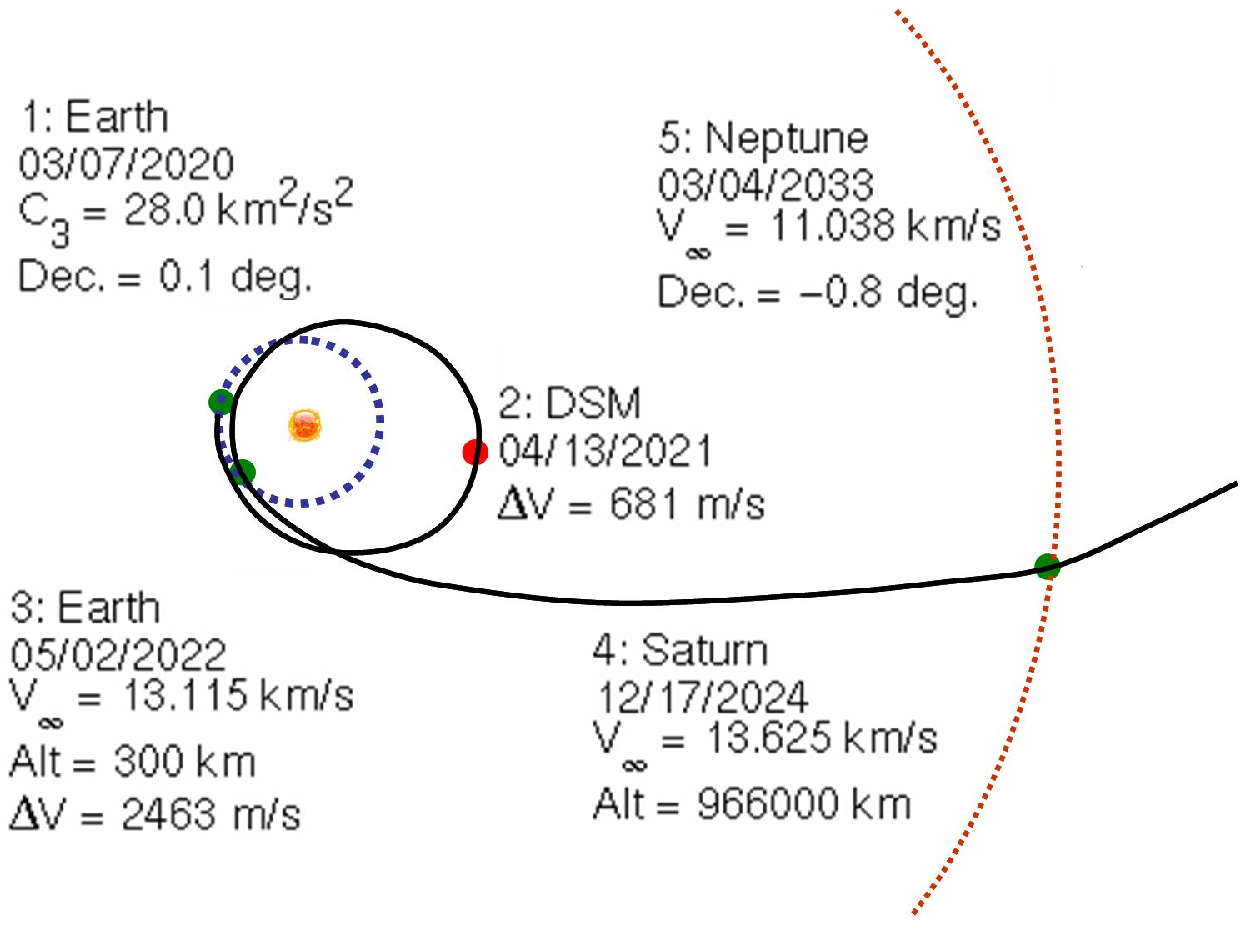} \\
\end{tabular}
\caption{OSS orbit for a launch in 2020, with Venus and 2 Earth gravity assists (VEEGA) on left and 2 Earth and Saturn gravity assists (EESGA) on right}
\label{fig:Orb}       
\end{figure}

For the last requirements different strategies have been analyzed.
A direct trajectory would enable almost annual launch windows at the expense of a relatively heavy launcher due to the high initial velocity required. Transfers using inner solar system gravity assists would allow less heavy launchers. Two optimized trajectories are compared in Figure \ref{fig:Orb}. 

Table \ref{tab:Launch} gives the main characteristics of the different orbit strategies and the delivered mass to Neptune transfer orbit.
C3 is the square of the hyperbolic excess velocity (V$\infty$). Table gives C3 with respect to Earth at departure and V$\infty$ at Neptune arrival.
For the direct launch, Star48 propulsion module of Atlas 551 launcher is required. For trajectories with gravity assistance, 
the $\Delta$V presented doesn't take into account orbit control which requires a minimal $\Delta$V of 120~m/s (50~m/s for the launch window, 30~m/s for the launcher dispersion correction, 30 m/s for trajectory correction during cruise and 10~m/s for planet approach), leading to 34~kg of mono-propellant and inert mass for a spacecraft of 500~kg. 

\begin{table}
\caption{Orbit characteristics and delivered mass (kg) on Neptune transfer orbit}
\label{tab:Launch}       
\begin{tabular}{|c|c|c|c|c|c|c|c|c|}
\hline
Orbit & C3 (Earth)     & $\Delta$V  & V$\infty$ (Neptune)   & Soyuz  & Atlas  & Atlas \\
      & (km$^2$/s$^2$) & (m/s)      & (km/s)      & Fregat & 401    & 551 \\
\hline
Direct& 159.5          &    -       & 8.48        & -      & -      & 500 \\
\hline
VEEGA & 14.7           & 3505       & 13.79       & 211    & 509    & 961 \\
\hline
EESGA  & 28.0           & 3144       & 11.04      & 164    & 452    & 890 \\
\hline
\end{tabular}
\end{table}

During Neptune and Triton encounter, the flyby velocity is less than the one for New Horizons for Pluto-Charon encounter, with the same type of instrumentation \citep{Guo-2008} and scientific objectives. The time allocation between the difference experiments should be analyzed during further steps.

\subsection{Spacecraft}\label{section:spacecraft}

The spacecraft design is based on the hypothesis of a direct launch, with 34~kg of propellant for orbit control. 
In case of an indirect launch, the additional $\Delta$V will have an impact on the size of the tank, or requires a specific propulsion module. 
The objective of this spacecraft preliminary design is to analyse the capability to carry all the type of instruments, even if the limited payload mass will need  to make a choice between them in a further step. 

\begin{figure}
  \includegraphics[width = 0.8 \linewidth]{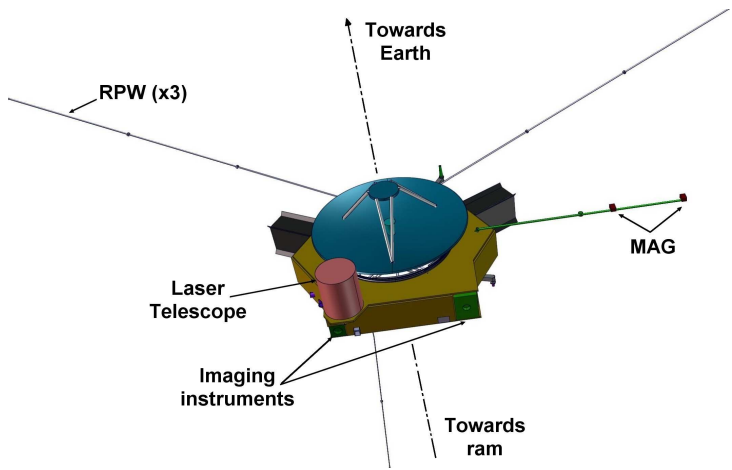}
\caption{Spacecraft as viewed from the terminator of the planetary object: accommodating the dust analyzer on the ram face and the RPW and the MAG on large booms protects their measurements from disturbances; the laser, which will have measured the Eddington parameter when closer to the Sun, has its boresight aligned with the HGA}
\label{fig:OSS_SC1}       
\end{figure}

\begin{figure}
  \includegraphics[width = 0.8 \linewidth]{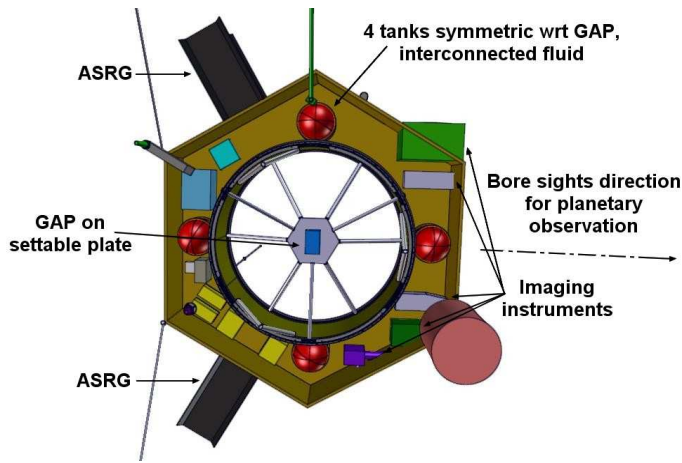}
\caption{Spacecraft viewed from top, with HGA and closure panels removed: the New Horizons-like planetary science payload accommodation ensures the planetary science objective while the distribution of the matter on a ring favours the accuracy of the measurements by the GAP.}
\label{fig:OSS_SC2}       
\end{figure}

The spacecraft architecture as illustrated in Figure \ref{fig:OSS_SC1} and \ref{fig:OSS_SC2} would allow to:
\begin{itemize}
	\item Provide a planet-pointing side with the observation instruments
	\item Accommodate the RPW and the dust analyzer properly with respect to velocity vector
	\item Accommodate the MAG with an adequate boom
	\item Accommodate the laser instrument for the measurement of the Eddington parameter with a pointing towards the Earth
	\item Provide the lowest and most axisymmetrical gravitational field as viewed from GAP
	\item Make coincide as much as possible the dry mass center of gravity, the propellant center of gravity, the radiation pressure force line and the GAP
	\item Ensure a stable and reliable alignment between the GAP and the high gain antenna (HGA) to ensure consistency between radio science and accelerometry
	\item Accommodate the two Advanced Stirling Radioisotope Generators (ASRG) required for the mission by minimizing their impact on the rest of the spacecraft (including radiation and asymetrical thermal emissions)
\end{itemize}

To this end the platform is built as a flat ring, exterior to a 1666~mm tube, easily interfaceable with Atlas~5 launcher, with the New-Horizons-like HGA on top, the GAP accelerometer at the center of the tube and at the center of gravity. The GAP is on a settable plate at the center, held by thermally stable struts.
Platform units are accommodated so as to balance the center of gravity.
 
The four Hydrazine tanks, a monopropellant with an Isp (220~s), are distributed symmetrically far from the GAP, with fluidic interconnection to balance the quantity of propellant. The knowledge of the propellant location is ensured by the membrane. In addition, a gauging method is proposed to limit the uncertainty on the quantity of propellant. The best compromise is a mix between book keeping offering a good precision at beginning of using and thermal gauging offering good performance at the end. From preliminary computation, the self-gravity uncertainty could be limited to 4~pm/s$^2$ along the antenna axis and 14~pm/s$^2$ along the transversal axis.

The two ASRGs are accommodated so that their radiators cannot couple with each other and the radiation they emit can be efficiently screened before it reaches the other units of the spacecraft. ASRG is a new type of Radioisotope Power System, and it is still being developed \citep{Chan-2007}. It uses a Stirling engine (with moving parts) to convert thermal energy to electricity with better efficiency than thermocouples. The output power is 146~W at the begin of lifetime and drops to 125~W after 14~years.

The upper face is generally pointed toward the Sun and Earth, hosting the HGA and the laser telescope. The downlink data rate will be similar to the one of New Horizons, but with the advantage of the Ka-band. With 70~m ground antenna, a data rate of 1~kbits/s could be achieved at a distance of 36~AU. Considering data size of 5~Gbits after compression as for New Horizons, 172 days are required to download the data using a session length of 8 hours per day. A storage is considered to store safely all the data before their transmission.

The lower face (launcher interface) is generally pointed towards the ram, bearing the dust analyzer and perpendicular to one of the three RPW.
One lateral "face" is pointable towards the planetary surfaces with all imaging instruments boresights in that direction.
\newline
\newline
Two modes are foreseen for the spacecraft: 3-axis stabilized mode during measurement and a spin-mode during the hibernation. The 3-axis stabilization allows high pointing accuracy during the entire measurement period. This mode is controlled by star trackers and gyro-stellar filter system and commanded by reaction wheels. The reaction wheels are desaturated with hydrazine thrusters. The same thrusters are used for re-orientation of the spacecraft or for switching in spin mode. During hibernation, the attitude control will switch to a spin stabilized Earth pointing mode, with an angular rate around 5 rpm, in order to limit ground operations and safe components lifetime on board.

The guidance and control system shall be capable of providing spin axis attitude knowledge of the spacecraft to better than $\pm 0.027^{\circ}$ (3$\sigma$) and spin phase angle knowledge within $\pm 0.30^{\circ}$ (3$\sigma$). The same knowledge is provided for all axes when the spacecraft is in 3-axis stabilized mode. 

A specific analysis has been done for the deep space gravity test. 
This test occurs during the interplanetary cruise with a global pointing accuracy requested of 1~mrad ($\pm 0.06^{\circ}$) of the accelerometer measurement axes with the inertial reference frame axes. 
This misalignment shall be shared between several contributors. 
An accuracy of 0.5 mrad~($\pm 0.03^{\circ}$) is allocated for the spacecraft pointing in accordance with its performance. 
During the deep space gravity measurement, the wheels unloading must be avoided: thrusters cannot avoid parasitic motions of the probe, saturating the accelerometer. Fortunately, in deep space, disruptive constant torque loading the wheels mainly comes from the solar flux pressure which is low far away and which acts if the probe is not axisymmetric around its sun axis (at 1 AU, the torque will be less than $4.6 10^{-6}$ Nm). Far from the sun it can be noticed that the sun - probe - earth angle is relatively small implying that the 2.1~m HGA is quasi pointed toward the sun offering a frontal circular surface to the sun avoiding torque. Implementing classical 12~Nms reaction wheels must allow avoiding unloading operation during more than one month.
\newline
\newline
In the current spacecraft design, the payload mass is only 48~kg, which does not allow to carry all the instruments proposed in section \ref{section:payload}. A choice shall be done in further steps.
The propellant budget is assessed at 34~kg, taking into account only the interplanetary cruise. The total mass is 540~kg (see Table \ref{tab:OSS_mass}), slightly higher than the capability for a direct launch, but inline with indirect launch with Atlas 551.

\begin{table}
\caption{OSS preliminary mass budget}
\label{tab:OSS_mass}       
\begin{tabular}{|l|c|}
\hline
 	&  Estimate in kg \\
\hline
Structure & 68 \\
\hline
Thermal Control & 21 \\
\hline
Mechanisms & 9 \\
\hline
Communications & 35 \\
\hline
Data Handling & 17 \\
\hline
AOCS & 31 \\
\hline
Propulsion & 16 \\
\hline
Power & 86 \\
\hline
Harness & 24 \\
\hline
Instruments & 48 \\
\hline
Propellant & 34 \\
\hline
Total mass estimate & 389 \\
\hline
Plus maturity margins & 450 \\
\hline
Plus system margin 20\% & 540 \\
\hline
\end{tabular}
\end{table}

The power is sized assuming two ASRGs, with an end of life capacity of 125 useful Watts each, hence 250~W of available power resource, which is comparable to that of the New Horizons spacecraft.
Given the higher payload consumption, the power subsystem is complemented by a 25~kg battery, recharged outside of fly-bys and of communication periods. The battery supplies the complement needed during fly-bys so as to allow the simultaneous operation of all instruments and radio science. This complement of 2400~W.h  at 70\% of depth of discharge will allow an increase of 100~W in the resources over a period of 24~h, so from 8~h before Triton fly-by to about 8~h after Neptune fly-by.

\section{Conclusions}\label{section:conclusion}

The main fundamental physics objective is the deep space gravity test driven by the research of deviations of the General Relativity in order to merge the four fundamental forces. In a context dominated by the quest for the nature of dark matter and energy, testing gravity at the largest scales reachable by man-made instruments is essential to bridge the gap with astrophysical and cosmological observations.
For that purpose, it is necessary to have the best accurate navigation of the probe during the interplanetary trajectory, achieved with by the high accuracy of the accelerometer GAP and highly precise observations of the position of OSS probe, owing to laser or radio science measurements. The request of high accuracy leads to major constraints on the spacecraft design due to the high accuracy of the main instrument GAP which measures all the vibrations and forces acting on the spacecraft at a level of some pm/s$^2$.
But, with respect to a typical outer planet flyby mission, the fundamental physics tests allows to have a scientific return largely before the arrival to Neptune which occurs 13 years after launch, without concurrence with the planetary instruments in terms of consumption or data rate.

The constraints on mass and cost of the mission imposed by ESA lead to the choice of one flyby of Neptune and Triton instead of ending up in an orbit around Triton. Nevertheless, this choice  allows to pursuit the mission with a flyby of a Kuiper Belt object, enabling to consolidate the hypothesis of the origin of Triton as an object captured by Neptune at the time of formation of the solar system. 
The interest to visit again Neptune and Triton after the Voyager 2 flyby, half a century ago, relies on ground-based observations of time-variable phenomena in the planet's atmosphere and the ring system. 
The scientific return of the OSS mission will fuel our knowledge of the farthest planet of the solar system owing to substantial technological progress made since the days of Voyager 2. Furthermore, the mission would very much benefit from the maturity of existing instruments, as the OSS planetary objectives can be addressed by an instrument suite similar to that flown on-board the New Horizons spacecraft that is presently underway to Pluto and its satellites.


\begin{acknowledgements}

The authors thanks the reviewers for their comments and corrections.

We gratefully acknowledge the support of the Argo Science team for supplying their Decadal Survey White Papers \citep{Spilker-2010, Hansen-2010a, Hansen-2010b, Stansberry-2010}.

This proposal was supported by CNES (France) through a phase 0 study managed by E. Hinglais.
\end{acknowledgements}



\bibliographystyle{spbasic}      



\end{document}